\documentstyle[epsf,psfig]{mn}

\newif\ifAMStwofonts

\ifoldfss
  \ifCUPmtlplainloaded \else
    \NewTextAlphabet{textbfit} {cmbxti10} {}
    \NewTextAlphabet{textbfss} {cmssbx10} {}
    \NewMathAlphabet{mathbfit} {cmbxti10} {} 
    \NewMathAlphabet{mathbfss} {cmssbx10} {} 
  \fi
  \ifAMStwofonts
    \ifCUPmtlplainloaded \else
      \NewSymbolFont{upmath} {eurm10}
      \NewSymbolFont{AMSa} {msam10}
      \NewMathSymbol{\upi}     {0}{upmath}{19}
      \NewMathSymbol{\umu}     {0}{upmath}{16}
      \NewMathSymbol{\upartial}{0}{upmath}{40}
      \NewMathSymbol{\leqslant}{3}{AMSa}{36}
      \NewMathSymbol{\geqslant}{3}{AMSa}{3E}

      \let\leq=\leqslant \let\le=\leqslant
      \let\geq=\geqslant \let\ge=\geqslant
    \fi
  \fi
\fi 

\ifnfssone
  \newmathalphabet{\mathit}
  \addtoversion{normal}{\mathit}{cmr}{m}{it}
  \addtoversion{bold}{\mathit}{cmr}{bx}{it}
  \newmathalphabet{\mathbfit} 
  \addtoversion{normal}{\mathbfit}{cmr}{bx}{it}
  \addtoversion{bold}{\mathbfit}{cmr}{bx}{it}
  \newmathalphabet{\mathbfss} 
  \addtoversion{normal}{\mathbfss}{cmss}{bx}{n}
  \addtoversion{bold}{\mathbfss}{cmss}{bx}{n}
  \ifAMStwofonts
    \ifCUPmtlplainloaded \else
      \UseAMStwoboldmath
      \makeatletter
      \new@mathgroup\upmath@group
      \define@mathgroup\mv@normal\upmath@group{eur}{m}{n}
      \define@mathgroup\mv@bold\upmath@group{eur}{b}{n}
      \edef\UPM{\hexnumber\upmath@group}
      \new@mathgroup\amsa@group
      \define@mathgroup\mv@normal\amsa@group{msa}{m}{n}
      \define@mathgroup\mv@bold\amsa@group{msa}{m}{n}
      \edef\AMSa{\hexnumber\amsa@group}
      \makeatother
      \mathchardef\upi="0\UPM19
      \mathchardef\umu="0\UPM16
      \mathchardef\upartial="0\UPM40
      \mathchardef\leqslant="3\AMSa36
      \mathchardef\geqslant="3\AMSa3E

      \let\leq=\leqslant \let\le=\leqslant
      \let\geq=\geqslant \let\ge=\geqslant
    \fi
  \fi
\fi 

\ifnfsstwo
  \DeclareMathAlphabet{\mathbfit}{OT1}{cmr}{bx}{it}
  \SetMathAlphabet\mathbfit{bold}{OT1}{cmr}{bx}{it}
  \DeclareMathAlphabet{\mathbfss}{OT1}{cmss}{bx}{n}
  \SetMathAlphabet\mathbfss{bold}{OT1}{cmss}{bx}{n}
  \ifAMStwofonts
    \ifCUPmtlplainloaded \else
      \DeclareSymbolFont{UPM}{U}{eur}{m}{n}
      \SetSymbolFont{UPM}{bold}{U}{eur}{b}{n}
      \DeclareSymbolFont{AMSa}{U}{msa}{m}{n}
      \DeclareMathSymbol{\upi}{0}{UPM}{"19}
      \DeclareMathSymbol{\umu}{0}{UPM}{"16}
      \DeclareMathSymbol{\upartial}{0}{UPM}{"40}
      \DeclareMathSymbol{\leqslant}{3}{AMSa}{"36}
      \DeclareMathSymbol{\geqslant}{3}{AMSa}{"3E}

      \let\leq=\leqslant \let\le=\leqslant
      \let\geq=\geqslant \let\ge=\geqslant
    \fi
  \fi
\fi 

\ifCUPmtlplainloaded \else
  \ifAMStwofonts \else 
    \def\upi{\pi}
    \def\umu{\mu}
    \def\upartial{\partial}
  \fi
\fi

\def\xte{XTE J0111.2-7317}
\def\ha{$H\alpha$}
\def\hb{$H\beta$}
\def\ewha{EW$H\alpha$}
\def\ew{EW}
\def\hg{$H\gamma$}

\newcommand{\ergcms}{ergs cm$^{-2}$ s$^{-1}$}

\newcommand{\ergss}{ergs s$^{-1}$}

\title{XTE J0111.2-7317 : a nebula-embedded X-ray binary in the SMC}
\author[M.J. Coe et al.]
       {M. J.~Coe,$^1$, N.J. Haigh,$^1$, C.A. Wilson$^2$,I. Negueruela$^3$ \\
       $^1$Department of Physics and Astronomy, Southampton University, SO17 
1BJ, UK\\
       $^2$NASA, MSFC, Huntsville, AL35812, USA \\
	$^3$Dpto. de F\'{\i}sica, Ingenier\'{\i}a de Sistemas y Teor\'{\i}a de
la Se\~{n}al, Universidad de Alicante, Apdo. 99, E03080 Alicante, Spain }
\date{Accepted .
      Received ;
      in original form
      }

\pagerange{\pageref{firstpage}--\pageref{lastpage}}
\pubyear{2003}

\begin{document}

\bibliographystyle{plain}

\maketitle


\begin{abstract}

The observed characteristics of the nebulosity surrounding the SMC
High Mass X-ray Binary \xte~are examined in the context of three
possible nebular types: SNR, bowshock and HII region. Observational
evidence is presented which appears to support the interpretation that
the nebulosity surrounding \xte~is an HII region.  The source
therefore appears to be a normal SMC Be X-ray binary (BeXRB) 
embedded in a locally
enhanced ISM which it has photoionised to create an HII region.  This
is supported by observations of the X-ray outburst seen with BATSE and
RXTE in 1998-1999. It exhibited characteristics typical of a giant or
type II outburst in a BeXRB including large spin-up rates, $Lx\ge10^{38}$
erg/sq.cm-s, and a correlation between spin-up rate and pulsed
flux. However, the temporal profile of the outburst was unusual,
consisting of two similar intensity peaks, with the first peak of
shorter duration than the second.

\end{abstract}

\section{Introduction and background}

The X-ray transient \xte~was first detected by the Proportional
Counter Array (PCA), on the Rossi X-ray Timing Explorer (RXTE) 
X-ray observatory, on October
the 29th 1998 ~\cite{chakrabarty98a,chakrabarty98b} in the 2-10keV
band.

The observed temporal, flux and spectral characteristics are typical
of BeXRBs and the detection of pulsations at 31 seconds further
strengthens this hypothesis.  In addition the source
location is coincident with the SMC, the assumption of membership
permitting an absolute luminosity to be calculated. Whilst towards the
upper end of the typical flux distribution, peaking at approximately
$2\times10^{38}ergs^{-1}$, it is consistent with a giant (Type II)
outburst from a BeXRB.

The optical counterpart was identified by Israel et al (1999) and
subsequently optical and IR measurements by Coe, Haigh and Reig 2000
(henceforth referred to as CHR) confirmed the
optical counterpart to be a B0-B2 luminosity class III-V star showing
a strong IR excess. Surprisingly, however, their \ha~images of the
field revealed an extended region of \ha~emission surrounding the source. They
suggested it could be a surrounding SNR or wind bow shock.  Otherwise
\xte~is in many ways typical of the growing population of known SMC
BeXRBs. If a SNR the size of 3.1pc, using the diameter-to-age
relationship $D=0.9t^{2/5}$ ~\cite{clarkcaswell1976}, implies an age of
120yrs. A typical expansion velocity of 10000km/s ~\cite{chev1977}
for Type II SNe implies an age of $\sim300$years. Such a young BeXRB
would be of particular interest.

Throughout this work \xte~refers both to the X-ray source and to the
optical counterpart.

\section{Observations}

\subsection{Optical images}

Initial photometry was reported by CHR who carried out Johnson,
Str{\"o}mgren and IR measurements.  For this work further Wide Field
Imager (WFI) images of the field taken on 26 July 1999 
were extracted from the ESO
archive. The WFI is a focal reducer-type camera which is permanently
mounted at the Cassegrain focus of the 2.2-m MPG/ESO telescope at La
Silla, Chile. The detector is a mosaic of $4\times2~4096\times8192$
CCDs with 0.24\arcsec~pixels. These observations targeted a different
object, but because of the large 34\arcmin $\times$33\arcmin~field
\xte~was also imaged. Useful images in [SII], \ha, [OIII] and
continuum regions close to \ha~were de-archived. Preprocessing was
performed using {\footnotesize STARLINK} software.

\subsection{IR measurements}

The source was observed as part of the DENIS IR survey of the
Magellanic Clouds in Sept 1996. This IR catalogue shows I=15.27$\pm$0.02 and
J=15.19$\pm$0.10 measurements, but not $K_S$, because of the source's
faintness. However, a Service Observation of the field was carried out
using IRIS2 on the AAT during commisioning time. The resulting K band
image was calibrated using three DENIS sources in the field and an
accurate determination of K=14.79$\pm$0.03 was achieved. The resulting
J-K value of 0.4 is typical of other BeXRB systems in the SMC and
indicative of an excess IR flux over the stellar continuum for the
B0-B2 III-V star classified by CHR.


\subsection{Optical spectroscopy}

A blue spectrum was obtained at the ESO 1.52m telescope at La Silla,
Chile, on November the 3rd 1999. The Boller and Chivens spectrograph
was used with the no: 33 holographic grating, covering the range from
the classification region ($4000-5000\AA$) up to $\sim5900\AA$ with a
resolution of 1\AA ~per pixel. The slit was oriented in a NW-SE line to
encompass the brightest parts of the nebulosity.

CHR reported results from spectroscopy in which the only features seen
were strong \ha~ and \hb~ emission lines, as expected for a BeXRB. The
measured equivalent widths were $-27\pm$0.3\AA ~for \ha~ and
$-3.8\pm$0.2\AA ~for \hb. However, a considerable fraction of this
emission is contamination from the nebular emission which crosses the
stellar spectrum. Because the nebulosity is brightest close to the
star, the sky/background subtraction carried out as a standard
procedure in the spectral reduction does not remove such
contamination. The nebular [SII] lines at $6716\AA$ and $6731\AA$
contaminate the stellar spectrum in the same way. All published
equivalent widths to date (CHR and Covino et al. 2001) have been
skewed by this effect.

\begin{figure}
\begin{center}
\psfig{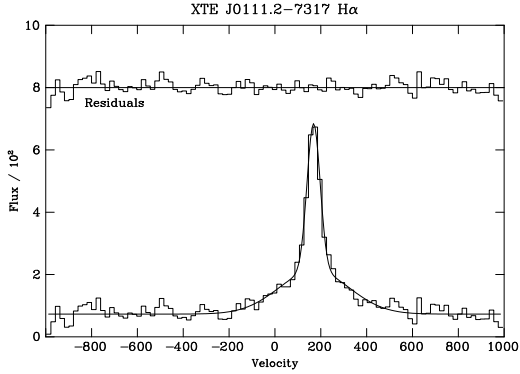}
\end{center}
\caption{\ha~region of SAAO \xte~spectrum with 2 component Gaussian fits.}
\label{fig:xtej0111saaospechalpha}
\end{figure}

Figure~\ref{fig:xtej0111saaospechalpha} shows the \ha~region of the
SAAO \xte~spectrum reported in CHR. Inspection of the spectral images shows that while the
nebular contamination is fully contained in a band 5 pixels wide
($\sim2.1\AA$ or 97km/s), the true circumstellar emission is
considerably broader at $\sim400$km/s, typical for Be stars. Thus
whilst the measured \ew is certainly an overestimate, there is no
doubt that intrinsic circumstellar emission exists. A two component
Gaussian fit to the line profile was performed within the DIPSO/ELF
package using the measured FWHM of the nebular contamination (obtained at a
spatial position adjacent to the star) to robustly remove it
(Figure~\ref{fig:xtej0111saaospechalpha}). This fit assigns $61\pm4\%$
of the line flux to the broader circumstellar disc (conventional Be)
component, producing an intrinsic \ewha ~of 16.5$\pm1.1$\AA.

The Doppler shift of the broad circumstellar component of \ha~is
recessional at 170.8$\pm9.3$km/s (visible in
Figure~\ref{fig:xtej0111saaospechalpha}) which agrees well with the
systemic value of 166$\pm$3km/s ~\cite{feast1961} for the SMC. Not only
does this confirm membership of the SMC (never really in doubt because
of the X-ray flux and photometry) but it also places a constraint on
the system's radial velocity relative to the local ISM of
$2.5\pm12.5$km/s. Accurate Doppler shifts for all of the lines
detected from the surrounding nebulosity are in
Table~\ref{tab:nebulalinelist}.

Regarding a classification based upon spectroscopic features, this has
been done by Covino et a. (2001) based upon superior spectra, and
their conclusion of B0.5-1Ve is consistent with this work.
 
\begin{figure}
\psfig{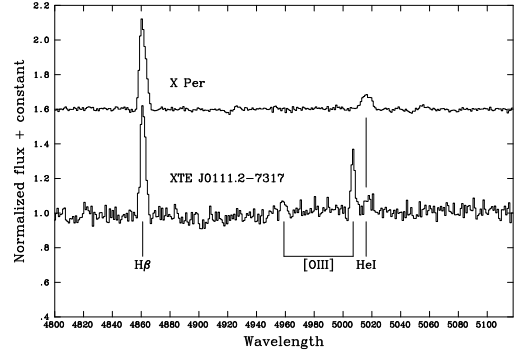}
\caption{The 4800--5050$\AA$ region of the ESO spectrum of \xte~showing \hb, [OIII] and HeI emission lines. The same range of a spectrum of the classic BeXRB X Per from the same observing run is shown for comparison. The \xte~spectrum has been shifted by $-165$km/s to the heliocentric rest frame.}\label{fig:OIII}
\end{figure}

Spectra (Figure \ref{fig:OIII}) show lines at 4959\AA~and 5007\AA,
attributed to [OIII] emission. Such forbidden line emission is typical
of many nebulae but uncharacteristic of the relatively dense
circumstellar environment of Be stars. A search through the spectra of
11 other LMC/SMC BeXRB spectra in the Southampton database, and
several galactic systems (including X Per and A0535+262) found no
other systems with features at this wavelength (as expected if it
arises from [OIII] emission). It is shown below that this feature
results from the superposition of nebular emission.

\subsection{X-ray data}

\subsubsection{Observations}

\begin{figure}
\psfig{file=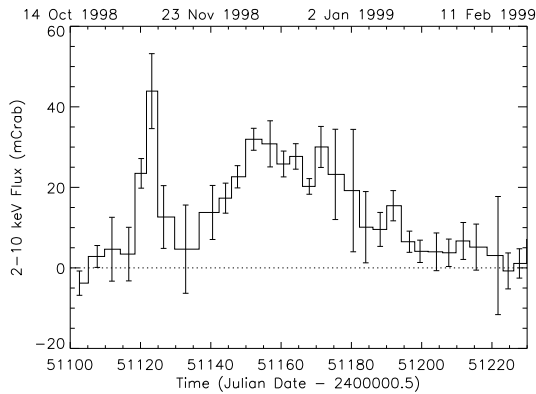,width=3in}
\caption{2-10 keV flux history for XTE J0111.2--7317 measured with the RXTE ASM.
All points are 4-day averages.
\label{fig:xr_asm}}
\end{figure}

XTE J0111.2--7317 was active in X-rays from 1998 October - 1999
February.  Figure~\ref{fig:xr_asm} shows the 2-10 keV flux history
measured with the All-Sky Monitor (ASM)~\cite{Levine96} on the RXTE.
XTE J0111.2--7317's location was in the field-of-view of the RXTE PCA
~\cite{Jahoda96} and High Energy Timing Experiment (HEXTE)
~\cite{Rothschild98} for 42 observations from 1998 October 20 - 1999
May 11. Further, the Burst and Transient Source Experiment (BATSE)
~\cite{Fishman89} on the Compton Gamma Ray Observatory had also
observed XTE J0111.2--7317's location from 1991-2000. No additional
outbursts have been detected with either BATSE or RXTE.

\subsubsection{Timing}

\begin{figure}
\psfig{file=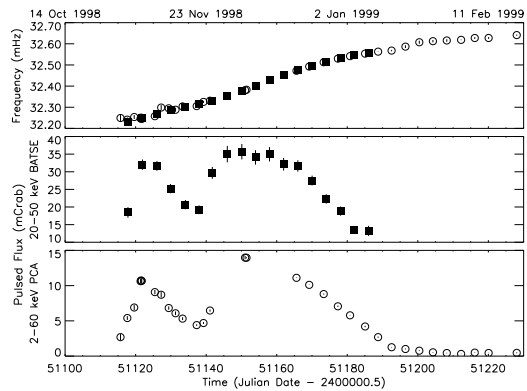,width=3in}
\caption{Top: Barycentered pulse frequencies measured with BATSE (filled
squares) at 4-day intervals and measured with the RXTE PCA for each observation
(open circles). Center: 20-50 keV pulsed flux measured with BATSE at 4-day
intervals. (1 mCrab = $1 \times 10^{-11}$ \ergcms)  Bottom: 2-60 keV pulsed flux
measured with the RXTE PCA for each observation. (1 mCrab = 2.786 cts s$^{-1}$
PCU$^{-1}$)
\label{fig:xr_hist}}
\end{figure}

To determine pulse frequencies for XTE J0111.2--7317, we performed a
grid search over a range of candidate frequencies using 1-second
resolution 20-50 keV data from the BATSE Large Area Detectors
(LADs). This technique is described in detail elsewhere 
~\cite{Finger99,Wilsonhodge99,Wilson02,Wilson03}. 
First we combined the count rates
over the 4 LADs viewing XTE J0111.2--7317, using weights optimized for
an exponential energy spectrum $f(E) = A \exp(-E/kT)$ with temperature
$kT = 12$ keV, and then grouped them into 300-s segments. In each
300-s segment, we fitted a model consisting of a sixth-order Fourier
expansion in pulse phase model (representing the 20-50 keV pulse
profile) and a spline function with quadratics in time (representing
the background). Using 300-s segments allowed use of a quadratic
background model, allowed us to effectively fit bright Earth
occultation steps in the data, improved computational efficiency, and
reduced smearing of the pulse profiles if the initial phase model was
incorrect.  The final step was to perform a grid search in frequency
using the set of typically several hundred estimated 20-50 keV pulse
profiles in each 4-day interval of data. For each grid point, the
segment pulse profiles were shifted according the the corresponding
frequency offset. These profiles were then combined into a mean
profile. The best frequency was selected using a modified $Z_n^2$
statistic, called $Y_n$ after ~\cite{Finger99}, given by $Y_n =
\sum_{h=1}^n |\bar \alpha_h|^2/\sigma_{\bar \alpha_h}^2$ where $\bar
\alpha_h$ is the mean Fourier coefficient for harmonic $h$ and
$\sigma_{\bar \alpha_h}^2$ is the sample variance from the segment
profiles.  Using sample variances from the segment profiles rather
than assuming Poisson statistics properly accounted for aperiodic
noise in XTE J0111.2-7317 and in any other sources in the large BATSE
field of view. A similar technique was used to generate pulse
frequency measurements for RXTE PCA observations from 2-60 keV
Standard 1 (125 ms, no energy resolution) data and from RXTE HEXTE
data.  The root-mean-squared (rms) pulsed fluxes were estimated at the
best fit frequency as $F_{\rm RMS} = (0.5 \sum_{h=1}^m |\bar
\alpha_h|^2)^{1/2}$.  Figure~\ref{fig:xr_hist} shows the barycentered
pulse frequency history measured with BATSE (squares) and with the
RXTE PCA (open circles), the 20-50 keV pulsed flux measured with
BATSE, and the 2-60 keV pulsed flux measured with the RXTE
PCA. Pulsations were detected with BATSE from 1998 October 30 - 1999
January 9. The RXTE PCA continued to detect pulsations until
observations ceased on 1999 February 19.  Pulsations were not detected
in two subsequent PCA observations on 1999 March 26 and 1999 May 11.

\begin{figure}
\psfig{file=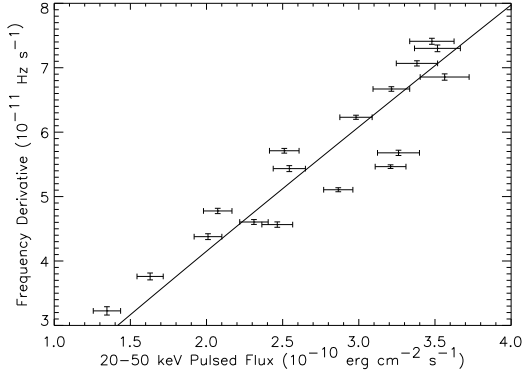,width=3in}
\caption{Frequency derivatives versus 20-50 keV pulsed flux
measured with BATSE. The solid line denotes a power law fitted to these data
with an index of $0.9 \pm 0.1$.
\label{fig:xr_fdotvsflux}}
\end{figure}

Pulse frequency derivatives were computed by differencing adjacent
BATSE pulse frequency measurements. Figure~\ref{fig:xr_fdotvsflux}
shows the correlation between frequency derivative and 20-50 keV
pulsed flux.  This correlation is well fitted with a power law with
index $0.9 \pm 0.1$, which is consistent with the index of $6/7$
expected from simple accretion theory, if accretion from a disk is
assumed. If a reliable bolometric correction can be derived for a
source, this correlation can be used to constrain the distance or the
magnetic field strength if the other is known. Unfortunately in this
case, large changes in the pulse fraction over the course of the
outburst prevented us from deriving a reliable bolometric flux (see
Figure 6).

\subsubsection{Energy Spectra}

Analysis of the energy spectrum of XTE J0111.2--7317 was complicated
by the fact that SMC X--1 was also present in the RXTE PCA and HEXTE
field-of-view in many of our observations. Using power spectra from
each observation, we determined in which observations 31-s pulsations
from XTE J0111.2--7317 were present and 0.7-s pulsations from SMC X--1
were not present. We then generated energy spectra and response
matrices for the PCA and HEXTE data for these 20 observations and
fitted them in XSPEC with an absorbed power-law with a high energy
cut-off and a Gaussian iron line. Average parameter values for the 20
observations are given in Table~\ref{tab:xr_spectra}. The flux in the
iron line was correlated with the total flux, with a correlation
coefficient of 0.97 and a chance probability of $2.5 \times 10^{-5}$,
suggesting that the iron line is intrinsic to XTE J0111.2--7317. No
other parameters were obviously correlated with flux nor did they show
clear evolution with time.

\begin{figure}
\psfig{file=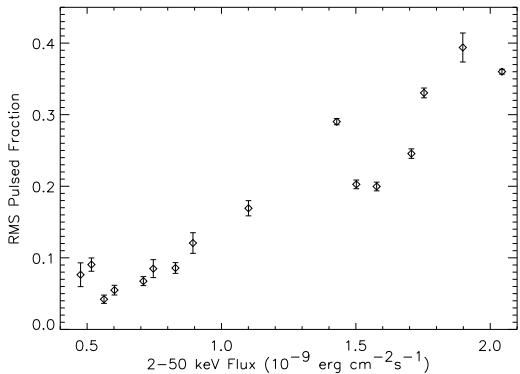,width=3in}
\caption{Root mean squared pulse fraction versus 2-50 keV flux for XTE
J0111.2--7317.
\label{fig:xr_pfrac}}
\end{figure}

Using only those observations where SMC X-1 was not present in the
power spectrum, we estimated pulse fractions from RXTE PCA data.
Figure~\ref{fig:xr_pfrac} shows the pulse fraction versus the 2-50 keV
flux computed from our XSPEC fits. The pulse fraction is correlated
with the total flux, indicating that the increase in flux is primarily
in the pulsed component.

\section{Discussion}

\subsection{The X-ray observations}

\begin{figure}
\psfig{file=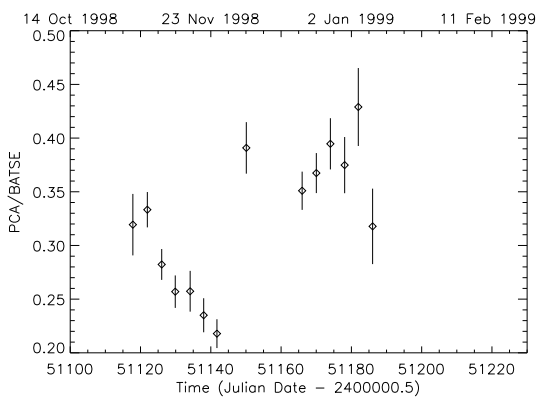,width=3in}
\caption{Ratio of the 2-60 keV pulsed flux in mCrab units measured with the RXTE
PCA to the 20-50 keV pulsed flux in mCrab units measured with BATSE.
\label{fig:xr_hr}}
\end{figure}

Comparing BATSE and PCA pulsed fluxes, we see evidence that the first
peak of the outburst is harder than the second peak, shown in
Figure~\ref{fig:xr_hr}.  Unfortunately, because SMC X-1 is present in
all of the observations during the first peak of the outburst, we
cannot study the effects of this change in hardness on the shape of
the energy spectrum in detail. The physical explanation of this is
unclear. We speculate that this may be an obscuration or absorption
effect - the softer X-rays are partially absorbed or blocked by the
large amount of material present at the beginning of the outburst. As
the outburst progresses, this material begins to dissipate and the
softer X-rays become more easily seen. A comparison of HEXTE and PCA
pulsed fluxes suggests a similar change in hardness, although the
HEXTE fluxes are much less certain than the BATSE fluxes due to the
smaller effective area and integration time. This suggests that the
effect is intrinsic to the source and is most likely not instrumental.

Observations of steady spin-up during the outburst allow us to compute
a lower limit to the luminosity of XTE J0111.2--7317. The angular
momentum of a rotating neutron star is given by $\ell = 2 \pi I \nu$
where $\nu$ is the spin frequency and $I$ is the moment of inertia of
the neutron star. The torque on the neutron star is obtained by
differentiating $\ell$ or $N = \dot \ell = 2 \pi I \dot \nu$. In an
accreting pulsar undergoing steady spin-up, the characteristic torque
is given by $N_{\rm max} = \dot M (G M_{\rm X} r_{\rm m})^{1/2}$ where
$\dot M$ is the mass accretion rate, $G$ is the gravitational
constant, $M_{\rm X}$ is the mass of the neutron star. and $r_{\rm m}$
is the magnetospheric radius. The maximum possible torque occurs when
the magnetospheric radius equals the corotation radius. Setting 
$N\leq N_{\rm max}$ yields an expression for $\dot M$ and hence
$L_{\rm X}$ that depends only on the spin frequency and its derivative
for assumed values of the neutron star parameters, i.e.
\begin{equation}
L_{\rm X} \geq (2 \pi)^{4/3} I (G M_{\rm X})^{1/3} R_{\rm X}^{-1} \dot \nu
\nu^{1/3}.
\end{equation}
Assuming $I = 10^{45}$ g cm$^{2}$, $M_{\rm X} = 1.4 M_{\odot}$, and $R_{\rm X}$
= 10 km, $L_{\rm X} \geq 10^{38}$ \ergss for typical observed values of $\nu
=  32.4$ mHz and $\dot \nu = 5 \times 10^{-11}$ Hz s$^{-1}$. The average 20-50
keV pulsed flux was $2.5 \times 10^{-10}$ \ergcms, implying that XTE
J0111.2--7317 is at a distance of $d \geq 59.8$ kpc, confirming that it is 
in the SMC. 

The outburst from XTE J0111.2--7317 should be classified as a giant or
Type II outburst given the large spin-up rates $(3-7) \times 10^{-11}$
Hz s$^{-1}$ measured with BATSE and the large luminosities $(2-9)
\times 10^{38}$ \ergss, measured in the PCA observations where SMC X-1
was not present. However, the temporal profile of this outburst is
unlike giant outbursts in other Be/X-ray binaries that typically
consist of either a single large intensity peak or a large peak
followed by smaller peaks modulated at the orbital period. The initial
large peak is usually much brighter and of much longer duration that
any succeeding peaks. XTE J0111.2--7317 instead has two peaks that are
similar in intensity with the first peak lasting about 25 days and the
second lasting about 60 days, followed by a declining tail lasting
about 30 days. The separation between the two peaks is about 30
days. This double peaked structure is most likely not due to orbital
effects because the frequency derivative is strongly correlated with
the pulsed flux and also shows the double peaked structure. Such a
correlation is predicted by simple accretion theory for systems
accreting from a disk. If the primary source of spin-up was orbital,
we would not expect a strong correlation with the pulsed flux and we
would expect to see both increases and decreases in pulse frequency as
the pulsar moved relative to the observer.

\begin{table}
\begin{center}
\caption{X-ray Spectral Parameters for XTE J0111.2--7317
\label{tab:xr_spectra}}
\begin{tabular}{l|ll} \hline
Parameter & Value & Standard Deviation \\ \hline
$N_{\rm H}$ (cm$^{-2}$)         & $1.5 \times 10^{22}$  & $0.5 \times 10^{22}$ \\
index                   &  0.93   & 0.06 \\
$E_{\rm cut}$ (keV)       & 13.4    &  0.8 \\
$E_{\rm fold}$ (keV)      & 20.3    &  2.2 \\
$E_{\rm line}$ (keV)      &  6.5    &  0.2 \\
$\sigma_{\rm line}$ (keV) &  0.8    &  0.5 \\ \hline
\end{tabular}
\end{center}
\end{table}


\subsection{The nebulosity}
\label{section:nebulosity}

The low surface brightness of the nebulosity, and the presence of
faint field stars superimposed upon it and nearby, combine to make a
morphological interpretation difficult. The \ha~image of \xte ~taken
from the ESO data shows the nebulosity clearly but it is unclear
whether some of the brighter patches are stars or enhanced nebular
emission. However, because the \ha~filter bandpass is effectively a
subset of the broader R filter, the stellar continuum emission present
in the \ha~images can be largely removed using an R band image.

The final result shows that the emission is smoother than would appear
from the \ha~image alone, though two slightly brighter features are
weakly visible to the SSW.

\subsubsection{Image characteristics}
\label{section:imageinterp}
 
Initial analyses of the nebula used CCD imaging from the SAAO 1.0m
provided the 'first look'. Though images were obtained in all filters,
only the \ha~image shows nebulosity. Though \hb~emission is present
(Table \ref{tab:nebulalinelist}), it is apparently too weak to
register in the image, due both to the large amount of continuum also
transmitted and to the poor response of the CCD in the blue.

\begin{figure}
\centerline{\psfig{file=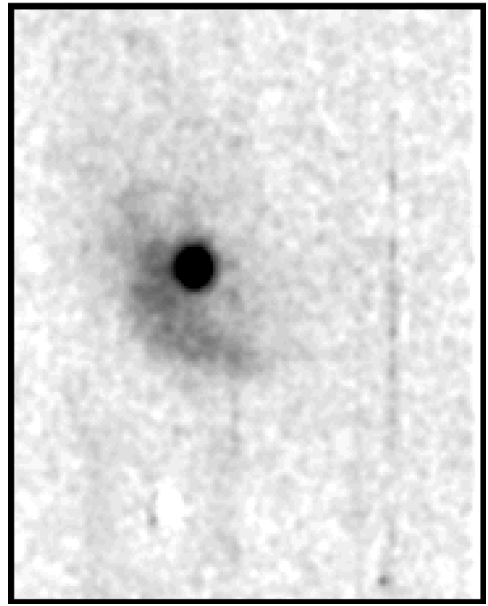,width=2.5in}}
\caption{Continuum subtracted \ha~image of the region around
\xte. Image from ESO La Silla 2.2m archive, smoothed with a
$\sigma=1\arcsec$ Gaussian filter.  Bright and dark lines are due to
bad CCD columns. Image size is $\sim$40 x 50 arcsec.}
\label{fig:xtej0111halpha}
\end{figure}

However these data were to some extent superseded by the archival images
from the ESO 2.2m telescope due to
their superior depth and resolution. Images in \ha, \ha~continuum,
[OIII] and [SII] were dearchived, but again only \ha~clearly displayed
nebulosity (Figure \ref{fig:xtej0111halpha}). Continua were removed
for both the [SII] and [OIII] images. Unfortunately only \ha~continuum
images were available to model the continuum component; this was not a
problem with [SII] (at $\lambda$6717,6731 c.f. \ha~at $\lambda6563$)
because of the minimal spectral separation, but the [OIII]
$\lambda5007$ continuum was not adequately represented by the
\ha~continuum image, thus leaving significant stellar images and
preventing the detection of the small [OIII] nebula (Section
\ref{section:linedistribution}). The FWHM of \xte~was measured both in
the un-subtracted and continuum-subtracted [OIII] images and found to
be consistent with the other field stars. [SII] appears to show
extremely weak emission coincident with \ha.

Thus analyses of the structure of the nebula in the light of these
species were undertaken spectroscopically (see below).Using the
maximum diameter $D$ of 20\arcsec~at a distance of 63.1pc one arrives
at a diameter of 6.1pc for the \ha~nebulosity.

Of the possible nebular types considered the simplest is a
conventional HII region, photoionised by \xte. A second possibility is
that the nebula is an SNR. In view of the association with the BeXRB
system, it would be extremely interesting to identify the SNR
corresponding to the formation of the system's neutron star. This
would, amongst other things, enable the system's age to be determined.

Another possibility which exists is that the nebulosity is a
bowshock. Such structures arise when ram pressure due to supersonic
motion of a star confines its expanding wind. The head or apex of the
structure defines a point where the momentum of the expanding stellar
wind balances that of the oncoming interstellar medium. Bowshocks are
commonplace around OB runaway stars. Indeed, according to the standard
theory of BeXRB formation, such systems MUST be runaways
~\cite{vandenheuvel2000}. Therefore bowshocks could be commonplace
around BeXRBs. The nature of the nebula is discussed below.


\subsubsection{Nebular spectrum extraction}
\label{section:bsspectrum}

No nebulosity had been observed when the SAAO spectra were taken so
the exposures were calculated to properly expose for the Be
star. However since the slit of the spectrograph was aligned E-W
subsequent examination of the raw spectral images revealed weak
nebular emission lines offset to one side of the stellar continuum.
This one-sidedness stems from the asymmetrical morphology of the
nebulosity.

Two spectra of the source were acquired, the second because the
intended star largely missed the slit on the first. However both
contained strong nebular features, and so were co-added. The exposures
were sequential and as a pair were bracketted by arc exposures; no
wavelength shift was detectable. All further reduction took place on
this co-added image. 

The SAAO blue exposure similarly shows only very weak nebular
\hb~emission. In view of the existence of a far higher S/N blue
spectrum from La Silla, no further work was undertaken on the blue
SAAO spectrum.

\begin{table*}
\centering
\begin{tabular}{|l|l|l|l|l|l|}

\hline
Species         & $\lambda$       & Rest $\lambda$ & Flux per arcsec    & Velocity   & FWHM       \\
                &  ($\AA$)        &  ($\AA$)       &($10^{-16}erg/cm^2/s$) 
&(km/s)&(km/s)\\
\hline
ESO blue        &&&&&\\
3600--5900$\AA$ &&&&&\\
$[$OII]         &$3728.12\pm0.14$ & 3726.16    &$25.3\pm3.7$  & $145.3\pm10.8$ &$176\pm20$\\
$[$OII]         &$3730.91\pm0.14$ & 3728.91    &$37.4\pm4.0$  & $148.4\pm10.8$ &$176\pm20$\\
H$\gamma$       &$4343.10\pm0.43$ & 4340.47    &$10.0\pm2.7$  & $169.7\pm29.4$ &$217\pm60$\\
\hb~            &$4864.31\pm0.10$ & 4861.33    &$17.7\pm1.3$  & $172.0\pm 6.0$ &$189\pm19$\\ 
$[$OIII]        &$4961.89\pm0.23$ & 4958.91    &$ 2.8\pm1.2$  & $168.1\pm13.6$ &$171\pm39$\\
$[$OIII]        &$5009.82\pm0.23$ & 5006.84    &$ 7.7\pm1.4$  & $166.4\pm13.5$ &$170\pm39$\\
\hline
SAAO red        &&&&&\\
6180--6930$\AA$ &&&&&\\
\ha~            &$6566.28\pm0.02$ & 6562.76    &$150.5\pm3.5$ & $168.3\pm8.3$ &$66\pm2$\\
$[$SII]         &$6720.03\pm0.06$ & 6716.47    &$25.1\pm3.5$  & $163.7\pm5.7$ &$47\pm6$\\
$[$SII]         &$6734.41\pm0.06$ & 6730.85    &$18.0\pm3.2$  & $163.4\pm5.7$ &$47\pm6$\\
\hline
\end{tabular} 
\caption{List of observed discrete emission lines from 
nebulosity around \xte. Wavelengths from Allen (1976), 
Meinel (1965) and Takami et al. (2001).}\label{tab:nebulalinelist}
\end{table*} 

Thus the {\footnotesize STARLINK} task {\footnotesize EXTRACT} was
used to extract the region immediately adjacent to the star including
these emission features; sky subtraction was from a region 10 pixels
below. Whilst sky subtraction was used to identify and reject telluric
and other non-nebular lines, actual measurement used subtraction of a
constant background to significantly reduce noise. This also enabled
telluric lines to be used as wavelength standards, for example the
3.682\AA~separation of telluric~\ha~from its SMC component provides a
highly accurate velocity. Care was taken to exclude any stellar
(including Be circumstellar disc) flux. Table~\ref{tab:nebulalinelist}
lists the nebular emission lines obtained from the SAAO red and ESO
blue spectra.

The ESO blue spectra taken on 3 November 1999 were extracted using the
same method.

Dividing the \ha~flux density integrated along the slit by the
approximate scale of 8\arcsec ~for the object one derives a \ha~flux
density of $\sim20\times10^{-16} erg/cm^2/s/arcsec^2$, which is close
to the value found for the bowshock associated with Vela X-1
~\cite{kaper97} of $\sim10\times10^{-16} erg/cm^2/s/arcsec^2$.

Because of the small number of counts from the nebular lines, the
extracted spectra are a spatial average over all the nebulosity which
fell upon the spectrograph slit. Without higher S/N data this is the
only way of extracting spectra of usable quality. Discussion on the spatial
distribution of the different emission lines is presented below.

\subsubsection{Line fitting}

\begin{figure}
\psfig{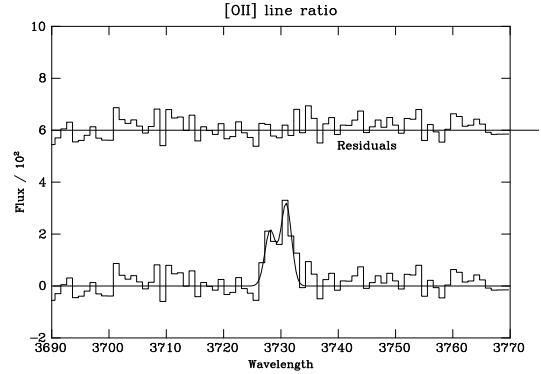}
\caption{Gaussian fit to the [OII] $\lambda$3727,3729 lines seen in the nebulosity around \xte.}
\label{fig:OIIfit}
\end{figure}

In order to derive the most accurate line fluxes the Emission Line
Fitting (ELF) suite within DIPSO was employed. 
The FWHM values were specified from
measurement of the \hb~line to reduce the number of free parameters
and thus errors. The final fit is shown in Figure~\ref{fig:OIIfit}.

The velocities are
derived from the shift relative to laboratory rest wavelengths, and
further corrected to a heliocentric reference frame using the
{\footnotesize STARLINK} program RV. The positive values refer to a
movement away from the observer, i.e. a redshift.

\subsubsection{Line flux calibration}

The presence of the stellar continuum alongside the nebular spectrum
allows an absolute flux calibration to be performed. Such calibration
is important for establishing reliable line ratios for diagnostic
techniques.

The reddened theoretical Kurucz model photospheric spectrum
~\cite{kurucz1979} fitted by CHR
was further reddened to allow for the
circumstellar reddening. This was normalized using the photometry at B
and R for the blue and red spectra respectively. The extracted stellar
spectrum was divided into this and fitted with a polynomial yielding
the required scaling factor at each wavelength to convert from counts
(the units of the line fitting) to $erg/cm^2/s/\AA$. Clearly this
calibration technique is dependent upon the Kurucz atmosphere being an
accurate model of the true stellar spectrum.
Note however that only the
shape of the spectrum within each spectral range is important; the
normalization, which is performed using photometry approximately in
the centre of each range, ensures that the flux density is correct at
this point and thus near the mark over the entire spectral range.

\subsection{Spectral results}

\subsubsection{Temperature.}
\label{section:temp}

The O line strengths constrain the nebular
electron temperature to below 25-40,000K; in fact standard nebular
theory provides a mechanism to maintain the temperature in the range
5000 - 10000K. Collisions of high energy electrons with ionised O,
Ne, S and Fe excite these ions into the metastable states which
subsequently decay by photon emission. This mechanism can be seen to
be at work from the high intensity of O and S forbidden lines. This
effectively limits the electron temperature to 10000K with a mean
value of 6700K ~\cite{kitchin}. The consistency with which this
mechanism has been seen to provide a 'thermostat' controling the
temperature of nebulae provides encouragement that it also applies in
this case.

\subsubsection{Nebular Density.}
\label{section:nebdensity}

[OII] and [SII] both emit close pairs of optical lines whose ratios
are diagnostic of $N_e$, the electron density ~\cite{saraph1970}. Using
ratios of lines produced by the same species is appealing
as uncertainties based on abundance, degree of ionization and the many
other variables cancel out. This diagnostic tool is based upon the
increasing effect of collisional de-excitation in preventing the
emission of a photon by an excited ion in a metastable state, as one
considers progressively higher densities.

In the neighbourhood of the critical densities from
$10^2-10^4cm^{-3}$ accurate densities may be derived, but outside of
this range the results are indistuinguishable from either 0 or
$\infty$ on each side of the critical density. The results in
Table~\ref{tab:lineratios} (derived from line fits shown in Figure
\ref{fig:OIIfit}) place an upper limit of
$N_e\le200cm^{-3}$.

\subsubsection{Line velocities.}
\label{section:linevel}

The heliocentric velocities of the nebular lines are all consistent
with a systemic SMC velocity of 166km/s, excepting the [OII]
$\lambda$3726,3729\AA. However, these lines lie at the extreme end of the
spectral range where the wavelength calibration is less accurate, so
this is not believed to represent a real velocity shift.

The FWHM are all consistent with instrumental broadening, with values
comparable to those of the calibration arc lines. The resolution of the
blue spectrum is such that structure with a velocity dispersion of up
to $\sim100$km/s could lie unresolved, whereas the red spectrum limits
such structure to $\sim30$km/s. This result argues strongly against a
SNR hypothesis for the nebulosity; even 'old' SNR display expansion
velocity dispersions of several hundred km/s ~\cite{snr1980}.

\subsubsection{Spatial distribution of nebular lines.}
\label{section:linedistribution}

The full known extent of the nebulosity is defined by the narrowband
\ha~image seen in Figure~\ref{fig:xtej0111halpha}. Whether this
represents the true distribution of material in the structure or is
more a reflection of the state of ionization can best be inferred by
looking at the distribution of emission from different atoms and
ionization states.

\begin{figure}
\psfig{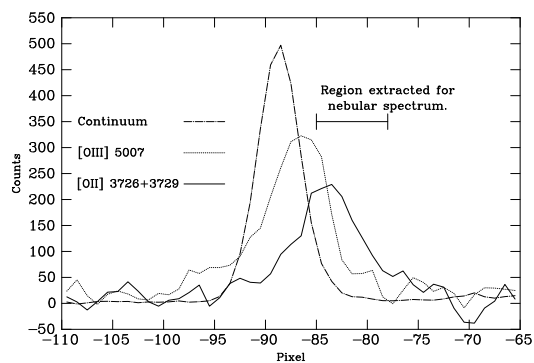}
\caption{Cross sections through the full widths of the [OII] and
[OIII] emission lines. The stellar contribution has been subtracted,
and is shown for reference. Profiles have been Gaussian smoothed with
$\sigma=0.7$.}
\label{fig:o2xsect}
\end{figure}

Representative cross-sections along the spatial axis of the spectral
images are shown in Figure 10, taken through the mid-point of the
detected emission lines with an appropriate width of pixels sampled
(mostly 3) depending upon the FWHM. These plots thus show the spatial
distribution of emission from each emission line; with the caveat that
the Be star produces a peak at its spatial location. This arises from
both photospheric emission and from the circumstellar disc.  These
peaks were removed by subtracting an extracted profile derived by
interpolation of the continua adjacent to each spectral line, in a
manner analogous to sky subtraction in conventional spectral
reduction. This was relatively successful for those lines which have
no photospheric or circumstellar counterpart:[SII], [O II] and [OIII],
though [O II] was affected by the rapid non-linearly declining
response of the CCD at short wavelengths.

In the case of the Balmer lines the combined spatial profile from the
photosphere and circumstellar disc has the same FWHM in the spatial
direction as the continuum, but must be scaled to account for the
circumstellar emission.

The orientation of the slit for the ESO blue spectrum - [OIII], [OII],
\hb~and \hg~data is NW - SE through the star and the brightest part of the
nebulosity, thus in the context of the bowshock scenario it
is usefully aligned along the direction of the standoff distance $l$.
The [OIII] line was expected as it is observed to be amongst the
strongest lines in many nebulae, including other bowshock nebulae, HII
regions and PNe.

Narrow band imaging of the field has not produced a detection at
[OIII] $\lambda$4959,5007 and in the
spectral data it is detected only weakly from the nebular regions
brightest in \hb. It is however observed as a strong feature
from regions much closer to the star itself, to the extent that
extraction of the stellar spectrum shows strong features at the
aforementioned wavelengths (Figure~\ref{fig:OIII}), whereas extraction
of the neighbouring nebular spectrum detects them only very
weakly. This can be understood with reference to
Figure 10 which shows
spatial profiles of the [OII] and [OIII] emission lines along the
spatial axis.

Whilst the [OIII] emission peaks close to the stellar location and rapidly
drops off, \hb~is observed to considerably larger distances. The
[OIII] emission profile is much more symmetric around the Be star,
with only minor enhancement in the direction of strongest Balmer
emission. Additionally [OIII] is seen from the NW side of the Be star
where there is negligible Balmer emission. Where \hb~and \hg~are
brightest there is scarcely any [OIII] emission at all.

The \hb, \hg~and [OII] emission appear
to be coincident, though few conclusions can be drawn from the weak
\hg~detection. All are seen only to the SE side of the Be star, in a
manner totally consistent with the distribution seen from the
\ha~image in Figure~\ref{fig:xtej0111halpha}. Detection from pixels
75.5 to 92.5 implies an angular diameter of 13.9\arcsec, or 4.3 pc.

The SAAO red spectrum - \ha~and [SII] - shows the distribution of
nebular \ha~emission. The size of the nebulosity appears to be the
same as that of \hb~and \hg, with detection out to slightly larger
radii because of the much stronger signal. [SII] appears to trace the
\ha~exactly.

\subsection{Nebular classification.}
\label{section:discussionneb}

If a spectral classification of B0.5-B1 (CHR and Covino et al, 2001) is
accepted it
becomes possible to calculate the Str{\"o}mgren radius within which a
uniform cloud of H is completely photoionized (note this does not
apply to the hollow bowshock model) and would be expected to emit
\ha. This radius is sensitive to spectral class around B0, in that
this classification sensitively determines the Lyman continuum flux
$Q_0$ depending upon the details of the stellar model used. 
Vacca et al. (1996)
tabulate $Q_0$ for O3-B0.5V stars, so though one must perform an
extrapolation to yield a result for B1, the function is a smooth one
and so the result expected to be accurate.
Incorporating the Martins et al. (2002) 
downward revision by 40\% of Vacca et
al.'s $Q_0$, one arrives at $4.8\times10^{47}$ at B0.5V and
$2.4\times10^{47}$ at B1V. 

Suitably modifying the derivation of
equation 6A.1.10 from Kitchin (1987), we arrive at the formula

\begin{equation}
R_S=\left( \frac{3Q_0}{4\pi 
N_e^{2}\times6.5\times10^{-15}T_e^{-0.85}}\right)^{\frac{1}{3}} 
(metres)
\end{equation}

Using the observed nebular radius of 3.05pc requires $N_e=7.6cm^{-3}$
at B0.5 or only $N_e=5.1cm^{-3}$ at B1V, at the lower end of observed
nebular densities, and in agreement with the results of Section
\ref{section:nebdensity} using forbidden line ratios. The use of SMC
metallicities, reducing line blanketting effects, raises $Q_0$ and
thus requires slightly higher densities, but this effect is only of
the order of a few \%.

This suggests the scenario that the nebulosity is simply an HII
region, perhaps part of a cloud adjacent to and ionized by \xte. If
so, and considering the fact that the calculated densities are very
much at the lower end of the range for HII regions, why is this
structure detectable at all? Most OB stars rid their neighbourhood of
gas early in their existence via powerful winds. The answer may lie in
the fact that as a BeXRB, \xte~has a large space motion
~\cite{vandenheuvel2000} and has impinged upon a cloud which it has
subsequently ionized. If this cloud increases in density towards the
south-east (lower left in the images), the slightly smaller radius in
this direction can be explained with the Str{\"o}mgren sphere argument
($R_S=fn(N_H)$), or as an ionization front propogating into the cloud.

\begin{table}
\centering
\begin{tabular}{|l|l|}
\hline
$[$OII] 3729:3726 & $1.48\pm0.27$\\
$[$SII] 6716:6731 & $1.40\pm0.31$\\
$[$SII]:\ha~      & $0.286\pm0.031$ \\
\hb:[OIII]       & $1.69\pm0.32$\\
\hline
\end{tabular} 
\caption{Diagnostic line ratios from nebulosity around \xte.}
\label{tab:lineratios}
\end{table} 

\subsubsection{The bowshock theory}

Stellar bow-shocks form from the
ram-pressure interaction of the local interstellar medium with the
stellar wind; when the star's velocity exceeds the sound velocity in
the ISM material is swept up into a shock front which is detected
primarily using one of two techniques: narrowband imaging (mostly \ha)
and IR imaging, mostly using IRAS data. Whereas \ha~imaging takes
advantage of excited H, the IRAS flux (usually most pronounced in the
$60\mu$m band) arises from thermal radiation from swept-up dust
originating from the star, heated by shock interactions.

The appearance of a bowshock is clearly heavily dependant upon the
angle that the star's velocity vector makes with respect to the plane
of the sky. Classic parabolic structures such as Vela X-1 and $\alpha$
Cam are seen when the star's motion lies
almost entirely within this plane. Less clearly defined structures are
more commonly seen, though enhanced brightness on one side, such as is
observed in \xte, is usually apparent ~\cite{noriega97}.

Thus the morphology of a bowshock is such that the head or apex of the
structure is well defined if viewed side-on, enabling a 'standoff
distance' $l$ to be determined; there is no sharply defined trailing
edge. The size of the structure is thus easily characterised only for
side-on specimens. 
In the case of \xte, the standoff
distance is approximately 3pc, within the range found for other
bowshocks tabulated in Table 4.

Comparing the $60\mu$m IRAS image of the $\alpha$ Cam bowshock in
with the \xte~nebula  reveals an extremely similar
appearance. Both nebulae lie within an approximately parabolic
perimeter, and though $\alpha$ Cam itself is not visible in the IRAS
image, both stars lie slightly inside the radius of curvature of the
nebulosity, agreeing with the theoretical radius of curvature for a
stellar wind bowshock of $(5/3)l$.

\begin{table*}
\centering
\begin{tabular}{|l|l|l|l|}
\hline
Object & Spectral class & Standoff dist.(pc) & Space vel.(km/s) \\
\hline
\xte~        & $B0Ve^1$            & 3.0     &                  \\
Vela X-1     & $B0.5Ibe^2$         & 0.48    & $\sim50km/s^8$     \\
$\alpha$ Cam & $O9.5Iae^3$         & 5.1$^6$ & 48$^7$           \\
0623+71      & $Cataclysmic Var.^4$& 0.08    & $\sim100km/s$    \\
Betelgeuse   & $M2Iab^5$           & 0.8     & 56               \\
\hline
\end{tabular} 
\caption{Sizes of some stellar bowshocks and spectral classification of their associated stars. 
Sources 1:This work, 2:Kaper et al. (1997), 3:Noriega-Crespo et
al. (1997), 4:Hollis et al. (1992), 5:Noriega-Crespo et al. (1997),
6:van Buren and McCray (1988), 7:Stone (1979), 8:Comeron \& Kaper (1998)}
\label{tab:bowshocks}
\end{table*}

Without doubt a measured space velocity for \xte~would be strong
evidence one way or the other. The radial velocity has been
investigated by cross correlating the blue ESO spectrum with that of a
velocity standard, but errors of 50km/s and a result consistent with
zero were sufficiently large to preclude any conclusions being drawn,
probably because of the low S/N of the spectrum. Studies of specific
lines would seem to suggest a limit on $v_r$ of less than 10km/s. 
The bowshock appears
however to be nearly edge on, in which case most of the stellar
velocity is oriented in the plane of the sky and a modest or zero
$v_r$ would be expected.

Comeron \& Kaper (1998) have performed numerical simulations of
bowshocks in a number of cases. One of their findings is that in low
velocity cases ($v_*\ll100$km/s), of which \xte~is an example as a
BeXRB, the bowshock becomes much thicker as the assumption of
instantaneous cooling, implicit in the above equation, fails. The
resulting thick layer of hot gas, bounded on one side by the stellar
wind, and on the other by the ISM, certainly can explain the broader
appearance of \xte's nebula compared to that of Vela X-1.

A second consequence is that peak density occurs at $1.5-2l$, somewhat
away than the balance point $l$ predicted by simple momentum
balance. Rearranging equation 16 of Raga et al. (1997) to incorporate
metallicity dependances ~\cite{vink2001} of $\dot M$ and $v_W$ gives
the following equation balancing the ram pressure of the ISM with that
of the expanding stellar wind at the standoff point, a distance $l$
from the star.

\begin{equation}
N_{ISM}=\frac{\dot M_{Z=1}Z^{0.82}v_W}{4\pi l^2m_Hv_*^2}
\end{equation}

Using a solar metallicity mass loss rate of $3\times10^{-8}M\sun
yr^{-1}$ ~\cite{vink2000}, a wind velocity of 1000km/s, the assumption
of a pure hydrogen ISM, system velocity $V_w=15km/s$ (mean BeXRB
velocity), and $l=3.0pc/(1.5-2)$ requires an ISM density of
$0.02-0.034 cm^{-3}$. This value is low for any ISM, particularly in a
gas rich galaxy like the SMC where densities of less than $0.1cm^{-3}$
are uncommon. The assumption of an unusually high wind speed and/or
mass loss rate can reconcile this situation, but both seem somewhat
unlikely.

Though this is clearly some way from an exact method of calculating
the ISM density, it seems safe to conclude that it must be low if we
believe the bowshock interpretation. We are therefore led to
questioning why this is the only one known around a BeXRB. The
requirement of supersonic velocity may be the reason; the typical ISM
sound speed of $v_S\alpha N^{1/2}\sim10km/s$ is only slightly below
the mean BeXRB velocity of 15km/s ~\cite{vandenheuvel2000}, so
perversely it could be only the BeXRB systems travelling through
relatively tenuous regions that are capable of producing
bowshocks. 

The argument in Section \ref{section:discussionneb} concerning the
Str{\"o}mgren radius around the B star does not apply to such thin
shells. Equation 21 of Comeron and Kaper (1998) considers the flow of
material into the bowshock to find the required ionizing flux - using
the values given above, the flux necessary to fully ionize the
material in the shock front is only $4\times10^{42} sec^{-1}$, which
is several orders of magnitude less than a B0.5-B1 star provides. Thus
we expect to see \ha~emission from the bowshock.

Whilst the [SII]:\ha~ratio of 0.286 is below the threshold of 0.4
usually accepted as clear evidence of shock excitation, it is
significantly above the typical values for photoionized nebulae which
mostly lie below 0.2. This suggests the possibility that a combination
of photoionization and shock excitation may be at work.

\subsubsection{Comparison with other bowshock spectra}
\label{section:comparisonspectra}

The discovery and study of bow-shocks around high velocity stars is a
relatively new facet of astronomy.
Few published spectra exist - spectra of only two
other stellar wind bowshocks have been found in the literature: Vela
X-1 ~\cite{kaper97} and a CV system 0623+71 ~\cite{hollis1992}. Both
sources show strong Balmer emission as well as [OIII] and [SII] lines,
in common with \xte. Additionally the 0623+71 spectra show strong
[OII] lines much like \xte. The only significant difference is the
absence of [NII] lines in the \xte~spectra, though the Vela X-1
spectra show them so weakly that if present in \xte~at a similar level
they would not have been detected. Thus the Vela X-1 nebular spectrum
is not significantly different to that of \xte, strengthening the
bowshock hypothesis.

Essentially the \xte~nebular spectra are qualitatively similar to both
of the bowshock spectra shown, in particular that of Vela X-1. This is
to be expected as the \xte~system far more closely resembles Vela X-1
than 0623+71.

The presence of [OIII] far from the presumed shock interaction
suggests photoionization as the mechanism for its excitation rather
than shock proceses; Noriega-Crespo and Garnavich (1992), in their
studies of bowshocks of Herbig-Haro objects, confirm earlier findings
~\cite{dopita84} that shock velocities of 80km/s are required to
provide sufficient excitation to produce [OIII] $\lambda$4959,5007, and
spectra do not reveal such velocities to be present. Bowshock theory
also does not predict this order of velocity.

\subsubsection{The HII region hypothesis.}

A high [OIII]:\hb~ratio is a diagnostic of photoionization from the Be
star's UV flux, comparable to the strong [OIII] lines seen in
planetary nebulae caused by the white dwarf's UV flux.The observed
ratio (Table \ref{tab:lineratios}) is somewhat higher than seen in the
two available bowshock spectra and is typical of HII nebulae and at the large
end of the distribution for SNR ~\cite{fesen1985,blair97}.

The observation of a small [OIII] nebula coincident with the central
star raises a possible problem for the bowshock theory. If this
material truly lies within the larger \ha~nebula, it would appear to
argue against the bowshock theory, as the interior of its
paraboloidal shell is considered to be empty, excepting the expanding
rarified stellar wind. The simulation work of Comeron \& Kaper (1998)
allows for some material to pervade this region, but probably not as
close to the star as we observe. Therefore in this scenario the [OIII]
emission must lie on the surface of this shell. In the case of \xte~however,
the chances of one patch of emission lying so well centred on our line
of sight to the star seem remote - the best explanation for this must
be that the [OIII] emission arises physically within the bounds of the
\ha~nebula, particularly as a Str{\"o}mgren sphere argument
can account for the relative sizes and locations of these features in
this scenario.

Section \ref{section:linedistribution} shows that the HII, [SII] and
[OII] nebulae are approximately coincident. Whereas this flows
naturally from a bowshock theory (one is simply measuring the standoff
distance $l$), it can also be understood in terms of Str{\"o}mgren
spheres. 

In view of the abundance of nebulae within 30\arcmin ~of \xte, the
association with a gas cloud does not cause any problems. SHASSA
(Southern H-Alpha Sky Survey Atlas) ~\cite{shassa} shows large HII
regions close by and diffuse emission to be extremely pervasive.

The high space velocity which all BeXRB possess could simply have
enabled the star to drift into an interstellar cloud and ionize it; it
could not have been in this position long as the surrounding gas would
have been blown away by the stellar wind. The mean space velocity of
BeXRBs of 15km/s ~\cite{vandenheuvel2000} corresponds to 1pc in 11,000
years.

\subsubsection{SNR hypothesis}
\label{section:rejection}

Obviously, the presence of the NS in a BeXRB requires there to have
been a SN explosion at some stage in the system's evolution. Thus the
discovery of a nebula surrounding the system naturally raises the
possibility that this may be a SNR, thereby enabling the age of the
system in its BEXRB state to be reliably estimated and providing
verification of the BeXRB formation mechanism. However, spectral
evidence shows that this is not the case.

Even a qualitative comparison of the nebular spectrum reveals that it
is much more closely allied both to published bowshock (discussed
above) and HII region
spectra than to a those of SNR. In addition, several quantitative
properties argue powerfully against it being a SNR.

The [SII]:\ha~ratio is a good distinguishing diagnostic between HII
regions and SNRs, and is the standard diagnostic used for classifying
nebulae in neighbouring galaxies where morphology cannot be used
~\cite{blair97}. With the exception of a handful of extremely low
surface brightness HII regions similar to and including the diffuse
interstellar gas, the vast majority of such objects have a
[SII]:\ha~ratio of less than 0.4, and usually below 0.2. SNR have
values normally significantly in excess of 0.5.
Thus with a ratio of 0.286 \xte~lies at the upper end
of the distribution of HII regions, suggesting the possibility of a
shock heating component in addition to photoionization as the dominant
mechanism.

The temperature sensitive line ratio [OIII] $\lambda5007$/[OII]
$\lambda3726+\lambda3729$ appears to be 0.12 from
Table~\ref{tab:nebulalinelist}, implying a temperature in excess of
70,000K (Figure 7 of Dopita, 1997) this is strongly affected by the
photoionized component at small distances from the star; no [OIII]
emission is detected from the shock-front itself implying temperatures
below those encountered in SNR. Similarly, whilst the [OIII]
$\lambda4363/\lambda5007$ ratio is traditionally used to provide
temperatures in SNR, upper limits to the $\lambda4363$ line
($3\times10^{-17}$ erg/s) combined with the [OIII] line strength can only
suggest temperatures below 25-40,000K, in the realm more typical of
planetary nebulae than SNR. The negligible presence of $O^{++}$ at the
shock location in itself betrays temperatures of below 40,000K
~\cite{dopita1977}. Additionally the lack of [O I] $\lambda$6300,6364
places the system well away from the realm of SNR in the diagnostic
diagrams of Fesen et al. (1985).

\subsection{Other HMXBs}

A total of 15 other HMXBs in the Magellanic Clouds were examined for
evidence of nebulosity, but none was found.  Anything down to a scale
of $\sim1$ pc and \ha~flux density $\sim20\%$ of the \xte~nebula
(which has almost exactly the same \ha~surface brightness as the Vela
X-1 bowshock) would have been detected. In view of the consistency of
the surface brightness of these structures 
and typical sizes it seems probable that any
other similar structures would have been found. Thus the frequency of
such objects in our sample is 1 in 15, or $6.7\%$. This statistic is
in exact agreement with the results of Huthoff and Kaper (2001) who
searched for IR bowshocks around 15 galactic HMXRBs, and found
none except for the well-known case of Vela X-1 ~\cite{kaper97}.

\section{Conclusions}

The observed characteristics of the nebulosity surrounding \xte~have
been examined in the context of three possible nebular types: SNR,
bowshock and HII region. Of these, the SNR hypothesis is excluded by
several line ratios and velocity dispersion. Such properties provide
no discrimination between bowshocks and HII regions, but the spatial
distribution of line emission from different species matches
predictions for an HII region to a high degree, whilst providing only
a poor match to the expectations for a bowshock. Overall these results
provide good evidence that the nebulosity surrounding \xte~is 
a conventional HII region.

The x-ray outburst from the 31-s X-ray pulsar in XTE J0111.2 observed
with BATSE and RXTE in 1998-1999 exhibited characteristics typical of
a giant or type II outburst in a BeXRB including large spin-up rates,
$Lx\ge10^{38}$ erg/sq-cm-s, and a correlation between spin-up rate and
pulsed flux. However, the temporal profile of the outburst was
unusual, consisting of two similar intensity peaks, with the first
peak of shorter duration than the second.

Thus, while apparently unusual, \xte~appears, in fact, to be a normal
SMC BeXRB embedded in a locally enhanced ISM which it has photoionised
to create an HII region.

\section*{Acknowledgments}

NJH acknowledges the use of STARLINK software and the support of a PPARC 
studentship. IN is partially supported by the Spanish Ministerio de
Ciencia y Tecnolog\'{\i}a under grants AYA2002-00814 and
ESP2002-04124-C03-03.


\end{document}